\documentclass[preprint, superscriptaddress,showpacs,preprintnumbers,amsmath,amssymb,pra]{revtex4}
\usepackage{graphicx}
\usepackage{amsmath}\usepackage{slashed}

\begin{document}

\thispagestyle{empty}

\title{Reflectivity properties of graphene with nonzero mass-gap parameter}

\author{
G.~L.~Klimchitskaya}
\affiliation{Central Astronomical Observatory at Pulkovo of the Russian Academy of Sciences,
Saint Petersburg,
196140, Russia}
\affiliation{Institute of Physics, Nanotechnology and
Telecommunications, Peter the Great Saint Petersburg
Polytechnic University, St.Petersburg, 195251, Russia}

\author{
V.~M.~Mostepanenko}
\affiliation{Central Astronomical Observatory at Pulkovo of the Russian Academy of Sciences,
Saint Petersburg,
196140, Russia}
\affiliation{Institute of Physics, Nanotechnology and
Telecommunications, Peter the Great Saint Petersburg
Polytechnic University, St.Petersburg, 195251, Russia}

\begin{abstract}
The reflectivity properties of graphene with nonzero mass-gap parameter are investigated
in the framework of Dirac model using the polarization tensor in (2+1)-dimensional
space-time. For this purpose, a more simple explicit representation for the polarization
tensor along the real frequency axis is found. The approximate analytic expressions for
the polarization tensor and for the reflectivities of graphene are obtained in different
frequency regions at any temperature. We show that the nonzero mass-gap parameter has a
profound effect on the reflectivity of graphene. Specifically, at zero temperature the
reflectivity of gapped graphene goes to zero with vanishing frequency. At nonzero
temperature the same reflectivities are equal to unity at zero frequency. We also find
the resonance behavior of the reflectivities of gapped graphene at both zero and nonzero
temperature at the border frequency determined by the width of the gap. At nonzero
temperature the reflectifities of graphene drop to zero in the vicinity of some frequency
smaller than the border frequency. Our analytic results are accompanied with numerical
computations performed over a wide frequency region. The developed formalism can be used
in devising nanoscale optical detectors, optoelectronic switches and in other optical
applications of graphene.
\end{abstract}

\maketitle

\section{Introduction}

Graphene is a two-dimensional sheet of carbon atoms arranged in a hexagonal
lattice, which possesses unusual mechanical, electrical and optical properties.
At energies below a few eV the electronic excitations (quasiparticles) in
graphene are well described by the Dirac model, i.e., display the linear
dispersion relation, where the speed of light $c$ is replaced with the Fermi
velocity $v_{F}\ll c$~\cite{1,2}. This makes graphene a unique laboratory for
testing some effects of quantum electrodynamics and quantum field theory, such
as the thermal Casimir force~\cite{3}, the Klein paradox~\cite{4}, and the
creation of particles from vacuum in strong external fields~\cite{5,6,7}. It
is important also that technological applications of graphene are numerous and
diverse. One could mention graphene coatings used in the optical detectors~
\cite{8}, solar cells~\cite{9}, transparent electrodes~\cite{10}, corrosium
protection~\cite{11}, optical biosensors~\cite{12}, optoelectronic switches
~\cite{13}, etc.

In order to make the best use of graphene in both fundamental physics and its
applications, it is necessary to investigate the reflectivity properties of
this novel material over the wide frequency region. As was shown experimentally,
in the region of visible light, where the photon energy is much larger than both
thermal and Fermi energies, the transparency and conductivity of graphene are
defined by the fine structure constant $\alpha=e^2/(\hbar c)$~\cite{14} and by
the quantity $e^2/(4\hbar)$~\cite{15}, respectively. These facts were understood
in theoretical studies of the conductivity of graphene using the current-current
correlation function and the Kubo formula (see Ref.~\cite{16} and review in
Ref.~\cite{17}).

The obtained results for the in-plane (longitudinal) conductivity have been
applied to investigate the reflectivity properties of graphene. This was done
in the framework of spatially local approximation, i.e., with no regard for
dependence on the wave vector. The reflection coefficients and reflectivities
of graphene with zero mass-gap parameter (quasiparticle mass) for the transverse
magnetic (TM), i.e., $p$ polarized, electromagnetic waves have been found at
both low~\cite{18} and high~\cite{19} frequencies. The case of gapless graphene
deposited on a substrate was considered in Ref.~\cite{20}.

The most fundamental quantity allowing full investigation of the reflectivity
properties of graphene over the entire frequency range is the polarization
tensor in (2+1)-dimensional space-time. It was found in Refs.~\cite{21,22} at
zero and nonzero temperature, respectively, and used to investigate the
Casimir effect in graphene systems in Refs.~\cite{23,24,25,26,27,28,29,30}
(several calculations of the Casimir force were performed also using the
density-density correlation functions of graphene and other methods
~\cite{3,31,32,33,34,35}). We stress, however, that for calculation of the
Casimir force using the Lifshitz theory one should use the reflection
coefficients defined at the discrete imaginary Matsubara frequencies
~\cite{36,37}. It was shown~\cite{38} that the polarization tensor of Ref.
~\cite{22} is valid strictly at the Matsubara frequencies and cannot be
immediately continued either to the real or to the entire imaginary frequency
axis.

The polarization tensor of graphene valid at all complex frequencies, including
the real frequency axis, was derived in Ref.~\cite{38}. At the imaginary
Matsubara frequencies it coincides with that of Ref.~\cite{22}, but, in
contrast to Ref.~\cite{22}, allows immediate analytic continuation to the
whole complex plane. Using this representation for the polarization tensor,
the reflectivity properties of gapless graphene were investigated at low, high
and intermediate frequencies for both polarizations of the electromagnetic
field TM and TE (i.e., transverse electric or $s$ polarization). In so doing,
some new properties were found which escaped notice in previous literature.
The polarization tensor of Ref.~\cite{38}, has been also used to find the
reflectivities of material plates coated with gapless graphene~\cite{39}.

In this paper, the polarization tensor of Ref.~\cite{38} is used to investigate
the reflectivity properties of graphene with nonzero mass-gap parameter $m$.
It is common knowledge that the quasiparticles of pristine (ideal) graphene
are massless. However, they acquire some small mass $m$ under the influence of
defects of the structure, electron-electron interaction, and in the presence
of substrates~\cite{1,40,41,42,43}. The exact value of $m$ for some specific
graphene sample usually remains unknown, but, according to some estimations
~\cite{21}, may achieve 0.1\,eV (we measure masses and frequencies in the units
of energy).
Here, we derive convenient explicit expressions
for the polarization tensor of graphene with nonzero mass-gap parameter at
any temperature. We demonstrate that under some conditions the nonzero mass-gap
parameter may have a dramatic effect on the reflectivity properties of graphene.
Specifically, we find the resonance behavior of graphene reflectivities at the
border frequency of incident electromagnetic waves
$\omega \approx 2mc^2/\hbar$. This effect
occurs at both zero and nonzero temperature.
At nonzero temperature the reflectivities of gapped graphene drop to zero at
some frequency smaller than the border one.
We find the asymptotic expressions
for the reflectivities of gapped graphene at both low and high frequencies. At
zero temperature the TM and TE reflectivities of graphene with $m \neq 0$ take
the zero value at zero frequency. This is different from the case of gapless
graphene, where both reflectivities at zero temperature are equal
to nonzero constants depending on $\alpha$ and on the angle of incidence~\cite{38}.
At nonzero temperature the reflectivities of gapped graphene go to unity with
vanishing frequency.
We also perform numerical computations of the reflectivities of gapped graphene
over the wide range of frequencies at both zero and nonzero temperature and find
the regions, where nonzero value of $m$ makes a profound effect on the obtained
results.

The paper is organized as follows. In Sec.~II we derive convenient expressions
for the polarization tensor of graphene with nonzero mass-gap parameter and
present the general formalism. Section~III contains calculation of the
reflectivities of gapped graphene at zero temperature. An investigation of the
impact of temperature on the reflectivity of gapped graphene is presented in
Sec.~IV. In Sec.~V the reader will find our conclusions and discussion.

\section{Special features of the polarization tensor of graphene with nonzero
mass-gap parameter}

The description of graphene by means of the polarization tensor in
(2+1)-dimensional space-time is in fact equivalent~\cite{26} to the method using the
density-density correlation functions, which is more often used in atomic and condensed
matter physics. The formalism of the polarization tensor, however, offers
some advantages over the more conventional methods, especially at nonzero
temperature, because it has been much studied and elaborated in the framework
of thermal quantum field theory.

It is conventional~\cite{44,45} to notate the polarization tensor as $\Pi_{\mu\nu}$,
where in (2+1)-dimensional case $\mu,\nu = 0, 1, 2$, and $\Pi_{tr}\equiv
\Pi_{\mu}^{\mu}$. In our configuration the polarization tensor $\Pi_{\mu\nu} =
\Pi_{\mu\nu}(\omega, \theta_i)$, where $\omega$ is the frequency of electromagnetic
wave incident on graphene and $\theta_i$ is the angle of incidence. We assume
that the graphene sheet is situated in the plane $(x,y)$ and the $z$ azis is
perpendicular to it. Then for real photons on a mass-shell the magnitude of the
wave vector component parallel to graphene (i.e., perpendicular to the $z$ axis)
is defined as
\begin{equation}
k_{\|}=\frac{\omega}{c}\sin\theta_i\,.
\label{eq1}
\end{equation}

The amplitude reflection coefficients for TM and TE polarizations of the
electromagnetic field incident on a graphene sheet are given by
~\cite{22,23,24,25,26,27,28,29,30,38}
\begin{eqnarray}
&&
r_{\rm TM}(\omega,\theta_i)=
\frac{\cos\theta_i \,\Pi_{00}(\omega,\theta_i)}{2i\hbar\frac{\omega}{c}\sin^2\theta_i
+\cos\theta_i \,\Pi_{00}(\omega,\theta_i)},
\nonumber \\[1mm]
&&
r_{\rm TE}(\omega,\theta_i)=
\frac{\Pi(\omega,\theta_i)}{2i\hbar\frac{\omega^3}{c^3}\sin^2\theta_i\,
\cos\theta_i -\Pi(\omega,\theta_i)},
\label{eq2}
\end{eqnarray}
\noindent
where the following notation is introduced:
\begin{equation}
\Pi(\omega,\theta_i)=\frac{\omega^2}{c^2}\left[\sin^2\theta_i\,
\Pi_{\rm tr}(\omega,\theta_i)+\cos^2\theta_i\,\Pi_{00}(\omega,\theta_i)\right]\,.
\label{eq3}
\end{equation}

Note that so important quantity as the conductivity of graphene can be
immediately expressed via the polarization tensor. Thus, the spatially nonlocal
(i.e., depending on the wave vector) longitudinal and transverse conductivities
of graphene are given by~\cite{26}
\begin{equation}
\sigma_{\|}(\omega,k_{\|})=-i\frac{\omega}{4\pi\hbar k_{\|}^2}\,\Pi_{00}(\omega,k_{\|}),
\qquad
\sigma_{\bot}(\omega,k_{\|})=i\frac{c^2}{4\pi\hbar k_{\|}^2\omega}\,\Pi(\omega,k_{\|}).
\label{eq4}
\end{equation}

It is convenient to represent the quantities $\Pi_{00}$ and $\Pi$ in the form
\begin{eqnarray}
&&
\Pi_{00}(\omega,\theta_i)=\Pi_{00}^{(0)}(\omega,\theta_i)+
\Delta_T\Pi_{00}(\omega,\theta_i),
\nonumber \\
&&
\Pi(\omega,\theta_i)=\Pi^{(0)}(\omega,\theta_i)+
\Delta_T\Pi(\omega,\theta_i),
\label{eq5}
\end{eqnarray}
\noindent
where $\Pi^{(0)}_{00}$ and $\Pi^{(0)}$ are defined at zero temperature and
$\Delta_T\Pi_{00}$, $\Delta_{T}\Pi$ are the thermal corrections which go to zero
in the limiting case of vanishing temperature.

The explicit expressions for the quantities $\Pi^{(0)}_{00}$ and $\Pi^{(0)}$ at
the imaginary frequencies in the case of arbitrary mass-gap parameter $m$ have
been found in Ref.~\cite{21}. In our case of real frequencies they can be
presented in the form
\begin{eqnarray}
&&
\Pi_{00}^{(0)}(\omega,\theta_i)=-\frac{\alpha\sin^2\theta_i}{\eta^2(\theta_i)}\,
\Phi(\omega,\theta_i),
\nonumber \\
&&
\Pi^{(0)}(\omega,\theta_i)=\alpha\frac{\omega^2}{c^2}\sin^2\theta_i
\Phi(\omega,\theta_i),
\label{eq6}
\end{eqnarray}
\noindent
where $\eta(\theta_i)\equiv\eta = [1 - \tilde{v}_{F}^2 \sin^{2}\theta_i]^{1/2}$,
$\tilde v_{F}\equiv v_F/c$ and the function
$\Phi$ is given by~\cite{38}
\begin{equation}
\Phi(\omega,\theta_i)=\left\{
\begin{array}{ll}
4mc-\frac{2\hbar\omega\eta}{c}\left[1+
\frac{4m^2c^4}{\hbar^2\omega^2\eta^2}\right]
{\rm arctanh}\frac{\hbar\omega\eta}{2mc^2},&
\hbar\omega<\frac{2mc^2}{\eta}, \\[1mm]
4mc-\frac{2\hbar\omega\eta}{c}\left[1+
\frac{4m^2c^4}{\hbar^2\omega^2\eta^2}\right]
\left({\rm arctanh}\frac{2mc^2}{\hbar\omega\eta}
+i\frac{\pi}{2}\right),&
\hbar\omega\geq\frac{2mc^2}{\eta}.
\end{array}
\right.
\label{eq7}
\end{equation}

Note that for a gapless graphene ($m = 0$) only the second line in Eq.~(\ref{eq7})
is applicable at all $\omega$. In this case the function $\Phi$ is pure imaginary
and has the negative imaginary part
\begin{equation}
\Phi(\omega,\theta_i)=-i\pi\hbar\frac{\omega}{c}\eta(\theta_i).
\label{eq8}
\end{equation}
\noindent
Substituting Eqs.~(\ref{eq6}) and ~(\ref{eq8}) in Eq.~(\ref{eq4}) with account of
Eq.~(\ref{eq1}), one finds the conductivities of graphene at zero temperature
\begin{equation}
\sigma_{\|}^{(0)}(\omega,k_{\|})=\frac{e^2}{4\hbar}\,
\frac{\omega}{\sqrt{\omega^2-\tilde{v}_F^2c^2k_{\|}^2}},
\qquad
\sigma_{\bot}^{(0)}(\omega,k_{\|})=\frac{e^2}{4\hbar}\,
\frac{\sqrt{\omega^2-\tilde{v}_F^2c^2k_{\|}^2}}{\omega},
\label{eq9}
\end{equation}
\noindent
which are in accordance with the previous knowledge (see, for instance,
Refs.~\cite{14,15,16,17,46}).

Now we consider thermal corrections to the 00 component of the
polarization tensor and to its trace entering the definition~(\ref{eq3}) of the
quantity $\Pi$. According to Eqs. (33)--(36) of Ref.~\cite{38}, along the real
frequency axis these thermal corrections can be written in the form
\begin{equation}
\Delta_T\Pi_{00({\rm tr})}(\omega,\theta_i)=\frac{16\alpha \hbar}{\tilde{v}_F^2}
\int_{0}^{\infty}\!\!\!dq\frac{q}{\Gamma(q)}\,\frac{1}{e^{\beta\Gamma(q)}+1}
\left[1+\frac{1}{2}\sum_{\lambda=\pm 1}
\frac{M_{00({\rm tr})}^{(\lambda)}(\omega,\theta_i,q)}{N^{(\lambda)}(\omega,\theta_i,q)}
\right],
\label{eq12}
\end{equation}
\noindent
where $\beta\equiv\hbar c/(k_BT)$.
Here, the integration variable $q$ has the dimension 1/cm,
$\Gamma(q)=[q^{2}+(mc/\hbar)^{2}]^{1/2}$, $k_B$ is the Boltzmann constant,
$T$ is the temperature, and the following notations are introduced:
\begin{eqnarray}
&&
M_{00}^{(\lambda)}(\omega,\theta_i,q)=4\Gamma^2(q)+\frac{\omega^2}{c^2}
\eta^2(\theta_i)+4\lambda\frac{\omega}{c}\Gamma(q),
\label{eq13}\\
&&
M_{{\rm tr}}^{(\lambda)}(\omega,\theta_i,q)=\frac{\omega^2}{c^2}
\eta^2(\theta_i)+4\tilde{v}_F^2\left(\frac{mc}{\hbar}\right)^2
+4\left(1-\tilde{v}_F^2\right)\left[\Gamma^2(q)+
\lambda\frac{\omega}{c}\Gamma(q)\right].
\nonumber
\end{eqnarray}
\noindent
The expression for $N^{(\lambda)}$ depends on the fulfilment of the following
conditions:
\begin{equation}
N^{(\lambda)}(\omega,\theta_i,q)=\left\{
\begin{array}{ll}
{\rm sign}Q_{\lambda}\left[{Q_{\lambda}}^{\!\!\!2}-\left(2
\tilde{v}_F\frac{\omega}{c}q\sin\theta_i\right)^2\right]^{1/2}, &
|Q_{\lambda}|\geq 2\tilde{v}_F\frac{\omega}{c}q\sin\theta_i, \\[1mm]
-i
\left[-{Q_{\lambda}}^{\!\!\!2}+\left(2\tilde{v}_F
\frac{\omega}{c}q\sin\theta_i\right)^2\right]^{1/2}, &
|Q_{\lambda}|< 2\tilde{v}_F\frac{\omega}{c}q\sin\theta_i,
\end{array}\right.
\label{eq14}
\end{equation}
\noindent
where
\begin{equation}
Q_{\lambda}\equiv Q_{\lambda}(\omega,\theta_i,q)=-
\frac{\omega^2}{c^2}\eta^2(\theta_i)-2\lambda\frac{\omega}{c}
\Gamma(q).
\label{eq15}
\end{equation}

In the case of $m \neq 0$ under consideration here it is more convenient to
represent the temperature correction ~(\ref{eq12}) as an integral with respect
to the new variable $u=\Gamma(q)$:
\begin{equation}
\Delta_T\Pi_{00({\rm tr})}(\omega,\theta_i)=\frac{16\alpha\hbar}{\tilde{v}_F^2}
\int_{mc/\hbar}^{\infty}\!\!\!du\frac{1}{e^{\beta u}+1}
\left[1+\frac{1}{2}\sum_{\lambda=\pm 1}
\frac{M_{00({\rm tr})}^{(\lambda)}(\omega,\theta_i,u)}{N^{(\lambda)}(\omega,\theta_i,u)}
\right],
\label{eq16}
\end{equation}
\noindent
where
\begin{eqnarray}
&&
M_{00}^{(\lambda)}(\omega,\theta_i,u)=4u^2+\frac{\omega^2}{c^2}
\eta^2(\theta_i)+4\lambda\frac{\omega}{c}u,
\label{eq17} \\
&&
M_{{\rm tr}}^{(\lambda)}(\omega,\theta_i,u)=\frac{\omega^2}{c^2}
\eta^2(\theta_i)+4\tilde{v}_F^2\left(\frac{mc}{\hbar}\right)^2
+4\left(1-\tilde{v}_F^2\right)\left(u^2+
\lambda\frac{\omega}{c}u\right).
\nonumber
\end{eqnarray}
\noindent
The quantity $Q_{\lambda}$ defined in Eq.~(\ref{eq15}) takes the form
\begin{equation}
Q_{\lambda}\equiv Q_{\lambda}(\omega,\theta_i,u)=-
\frac{\omega^2}{c^2}\eta^2(\theta_i)-2\lambda\frac{\omega}{c}u.
\label{eq18}
\end{equation}
\noindent
It is easily seen that for $\lambda=1$ the quantity $Q_{\lambda}$ is always
negative and satisfies the condition
\begin{equation}
|Q_{1}|\geq 2\tilde{v}_F\frac{\omega}{c}\left[u^2-
\left(\frac{mc}{\hbar}\right)^2\right]^{1/2}\sin\theta_i\,.
\label{eq19}
\end{equation}
\noindent
Thus, the first line in Eq.~(\ref{eq14}) is applicable leading to
\begin{equation}
N^{(1)}(\omega,\theta_i,u)=-\frac{\omega}{c}\left\{
\vphantom{\left[4\left(\frac{mc}{\hbar}\right)^2-
\frac{\omega^2}{c^2}\eta^2(\theta_i)\right]}
\eta^2(\theta_i)\left(
\frac{\omega}{c}+2u\right)^2
+\tilde{v}_F^2\sin^2\theta_i\left[4\left(\frac{mc}{\hbar}\right)^2-
\frac{\omega^2}{c^2}\eta^2(\theta_i)\right]\right\}^{1/2}.
\label{eq20}
\end{equation}

The case $\lambda=-1$ is more complicated and deserves separate consideration.
According to Eq.~(\ref{eq18}) in this case $Q_{-1}$ can be both positive and
negative. In order to explicitly rewrite Eq.~(\ref{eq14}) in terms of the
variable $u$, we introduce the quantity
\begin{eqnarray}
&&
\chi(\omega,\theta_i,u)\equiv {Q^{\,2}_{-1}}-4\tilde{v}_F^2\frac{\omega^2}{c^2}\left[u^2-
\left(\frac{mc}{\hbar}\right)^2\right]\sin^2\theta_i
\label{eq21} \\
&&~~=\frac{\omega^2}{c^2}\left\{
\vphantom{\left[4\left(\frac{mc}{\hbar}\right)^2-
\frac{\omega^2}{c^2}\eta^2(\theta_i)\right]}
\eta^2(\theta_i)\left(
\frac{\omega}{c}-2u\right)^2
+\tilde{v}_F^2\sin^2\theta_i\left[4\left(\frac{mc}{\hbar}\right)^2-
\frac{\omega^2}{c^2}\eta^2(\theta_i)\right]\right\}.
\nonumber
\end{eqnarray}
\noindent
The sign of the quantity $\chi$ under different values of parameters is
found by solving the equation
\begin{equation}
\chi(\omega,\theta_i,u)=0
\label{eq22}
\end{equation}
\noindent
with respect to the variable $u$:
\begin{equation}
u^{(\pm)}=\frac{1}{2}\left[\frac{\omega}{c}\pm\tilde{v}_F\sin\theta_i
\sqrt{\frac{\omega^2}{c^2}-\frac{4m^2c^2}{\hbar^2\eta^2(\theta_i)}}\right].
\label{eq23}
\end{equation}

If the condition
\begin{equation}
\omega<\frac{2mc^2}{\hbar\eta(\theta_i)}
\label{eq24}
\end{equation}
\noindent
is satisfied, Eq.~(\ref{eq22}) does not possess real roots and the function
$\chi$ is always positive. In this case the first line in Eq.~(\ref{eq14}) is
applicable and
\begin{equation}
N_{-1}(\omega,\theta_i,u)=\sqrt{\chi(\omega,\theta_i,u)},\quad
\frac{mc}{\hbar}\leq u<\infty,
\label{eq25}
\end{equation}
\noindent
where $\chi$ is defind in Eq.~(\ref{eq21}).

Now we consider the opposite condition
\begin{equation}
\omega\geq\frac{2mc^2}{\hbar\eta(\theta_i)},
\label{eq26}
\end{equation}
\noindent
when Eq.~(\ref{eq22}) has two different real roots $u^{(\pm)}$ defined in
Eq.~(\ref{eq23}) [they coincide in the case of equality in Eq.~(\ref{eq26})].
It is easily seen that under the condition~(\ref{eq26}) $u^{(-)} \geq mc/\hbar$.
Thus, in the regions $mc/\hbar \leq u \leq u^{(-)}$ and $u \geq u^{(+)}$ the function
$\chi$ is nonnegative and the quantity $N^{(-1)}$ is given by the first line of
Eq.~(\ref{eq14}). In so doing $Q_{-1}>0$ in the latter interval whereas it
can be both negative and positive in the former. As to the region
$u^{(-)}<u<u^{(+)}$, here the function $\chi$ is negative and, as a
consequence, $N^{(-1)}$ is given by the second line of Eq.~(\ref{eq14}).
Combining all these facts together, for the quantity $N^{(-1)}$ one obtains
\begin{equation}
N^{(-1)}(\omega,\theta_i,u)=\left\{
\begin{array}{ll}
{\rm sign}Q_{-1}\sqrt{\chi(\omega,\theta_i,u)},&
\frac{mc}{\hbar}\leq u\leq u^{(-)},\\
-i\sqrt{-\chi(\omega,\theta_i,u)},&
u^{(-)}<u< u^{(+)},\\
\sqrt{\chi(\omega,\theta_i,u)},&
u\geq u^{(+)}.
\end{array}
\right.
\label{eq27}
\end{equation}
\noindent
Simple analysis using Eqs.~(\ref{eq15}) and (\ref{eq23}) shows that
for $u\leq u^{(-)}$
within the interval
\begin{equation}
\frac{2mc^2}{\hbar\eta(\theta_i)}\leq\omega<\frac{2mc^2}{\hbar\eta^2(\theta_i)}
\label{eq28}
\end{equation}
\noindent
we have ${\rm sign}Q_{-1}=+1$,
whereas within the interval
\begin{equation}
\omega>\frac{2mc^2}{\hbar\eta^2(\theta_i)}
\label{eq29}
\end{equation}
\noindent
it holds ${\rm sign}Q_{-1}=-1$.

Equations (\ref{eq16})--(\ref{eq18}), (\ref{eq20}), (\ref{eq21}),
(\ref{eq25}) and (\ref{eq27}) are convenient for both analytic and numerical
calculations using the polarization tensor of gapped graphene. To apply them for
calculations of the reflection coefficients and reflectivities, it
is convenient also to present an explicit expression for the quantity
$\Delta_{T}\Pi(\omega,\theta_i)$. It is obtained from Eqs.~(\ref{eq3}),
(\ref{eq16}) and (\ref{eq17}) and has the form
\begin{equation}
\Delta_T\Pi(\omega,\theta_i)=\frac{16\alpha\hbar}{\tilde{v}_F^2}
\int_{mc/\hbar}^{\infty}\!\!\!du\frac{1}{e^{\beta u}+1}
\left[\frac{\omega^2}{c^2}+\frac{1}{2}\sum_{\lambda=\pm 1}
\frac{M^{(\lambda)}(\omega,\theta_i,u)}{N^{(\lambda)}(\omega,\theta_i,u)}
\right],
\label{eq30}
\end{equation}
\noindent
where the quantity $M^{(\lambda)}$ is given by
\begin{equation}
M^{(\lambda)}(\omega,\theta_i,u)=\frac{\omega^2}{c^2}\left[
\eta^2(\theta_i)\left(2u+\lambda\frac{\omega}{c}\right)^2
+
4\tilde{v}_F^2\left(\frac{mc}{\hbar}\right)^2\sin^2\theta_i\right].
\label{eq31}
\end{equation}
\noindent
The denominator $N^{(\lambda)}$ in Eq.~(\ref{eq30}) is presented in
Eqs.~(\ref{eq20}) and (\ref{eq27}).
In the limiting case $m \to 0$ the
above expressions take a more simple form used in Ref.~\cite{38} to investigate
the reflectivity properties of gapless graphene.

\section{Reflectivity of gapped graphene at zero temperature}

At $T=0$ the polarization tensor of graphene is given by Eqs.~(\ref{eq6})
and (\ref{eq7}). We consider first the region of frequencies (\ref{eq24}). Here,
the first line of Eq.~(\ref{eq7}) is applicable which can be rewritten in the form
\begin{equation}
\Phi(\omega,\theta_i)=
4mc\left[1-\frac{2mc^2}{\hbar\omega\eta}\left(1+
\frac{\hbar^2\omega^2\eta^2}{4m^2c^4}\right)
{\rm arctanh}\frac{\hbar\omega\eta}{2mc^2}\right].
\label{eq32}
\end{equation}
\noindent
This is real function in contrast with the case of gapless graphene [in the latter
case the function $\Phi$ is pure imaginary, see Eq.~(\ref{eq8})].

Substituting Eq.~(\ref{eq6}) with the real function $\Phi$ in Eq.~(\ref{eq2}),
one obtains the reflection coefficients and reflectivities of gapped graphene
at sufficiently low frequencies
\begin{eqnarray}
&&
{\cal R}_{\rm TM}(\omega,\theta_i)\equiv|r_{\rm TM}(\omega,\theta_i)|^2=
\frac{\alpha^2\cos^2(\theta_i)\Phi^2(\omega,\theta_i)}{4\hbar^2
\frac{\omega^2}{c^2}\eta^4(\theta_i)+\alpha^2\cos^2(\theta_i)\Phi^2(\omega,\theta_i)},
\nonumber \\
&&
{\cal R}_{\rm TE}(\omega,\theta_i)\equiv|r_{\rm TE}(\omega,\theta_i)|^2=
\frac{\alpha^2\Phi^2(\omega,\theta_i)}{4\hbar^2
\frac{\omega^2}{c^2}\cos^2(\theta_i)+\alpha^2\Phi^2(\omega,\theta_i)},
\label{eq33}
\end{eqnarray}
\noindent
where $\Phi$ is defined by Eq.~(\ref{eq32}). It is seen that at the normal
incidence ($\theta_i=0$)
\begin{equation}
{\cal R}_{\rm TM}(\omega,0)={\cal R}_{\rm TE}(\omega,0),
\label{eq34}
\end{equation}
\noindent
as it should be.

\subsection{Asymptotic results}

Now we consider the asymptotic regime of very small frequencies satisfying the
condition
\begin{equation}
\frac{\hbar\omega\eta(\theta_i)}{2mc^2}\ll 1.
\label{eq35}
\end{equation}
\noindent
Expanding Eq.~(\ref{eq32}) in powers of small parameter (\ref{eq35}), one obtains
\begin{equation}
\Phi(\omega,\theta_i)\approx
-\frac{4\hbar^2\omega^2\eta^2(\theta_i)}{3mc^3}.
\label{eq36}
\end{equation}
\noindent
Substituting this in Eq.~(\ref{eq33}) with account of Eq.~(\ref{eq35}), we
arrive at
\begin{equation}
{\cal R}_{\rm TM}(\omega,\theta_i)\approx \frac{4}{9}
\alpha^2\cos^2(\theta_i)\frac{\hbar^2\omega^2}{m^2c^4},
\qquad
{\cal R}_{\rm TE}(\omega,\theta_i)\approx \frac{4
\alpha^2\hbar^2\omega^2}{9m^2c^4\cos^2(\theta_i)}.
\label{eq37}
\end{equation}
\noindent
Note that in  Eq.~(\ref{eq37}) and below we have also used
$\eta(\theta_i)\approx 1$.

As can be seen in Eq.~(\ref{eq37}),
\begin{equation}
{\cal R}_{\rm TM}(\omega,\theta_i),{\ }{\cal R}_{\rm TE}(\omega,\theta_i)
\to 0{\ \ }\mbox{when}{\ } \omega\to 0.
\label{eq38}
\end{equation}
\noindent
This is different from the case of gapless graphene at zero temperature. In the
latter case, substituting Eq.~(\ref{eq8}) in Eq.~(\ref{eq6}), one finds
\begin{equation}
\Pi_{00}^{(0)}(\omega,\theta_i)=i\alpha\pi\hbar\frac{\omega}{c}\,
\sin^2\theta_i,
\qquad
\Pi^{(0)}(\omega,\theta_i)=-i\alpha\pi\hbar\frac{\omega^3}{c^3}\,
{\sin^2\theta_i}.
\label{eq39}
\end{equation}
\noindent
Substituting these equations in Eq.~(\ref{eq2}), after simple calculation,
we obtain
\begin{equation}
{\cal R}_{\rm TM}(\theta_i)=
\frac{\alpha^2\pi^2\cos^2\theta_i}{(\alpha\pi\cos\theta_i+2)^2},
\qquad
{\cal R}_{\rm TE}(\theta_i)=
\frac{\alpha^2\pi^2}{(2\cos\theta_i+\alpha\pi)^2}.
\label{eq40}
\end{equation}
\noindent
Note that both reflectivities in Eq.~(\ref{eq40}) do not depend on the frequency
and, thus, take nonzero values at $\omega=0$, as opposed to Eq.~(\ref{eq38}) for
a gapped graphene.

We are coming now to the asymptotic region
\begin{equation}
\omega\to\frac{2mc^2}{\hbar\eta(\theta_i)},
\label{eq41}
\end{equation}
\noindent
where the quantity on the left-hand side of Eq.~(\ref{eq35}) approaches unity
from the left. It is convenient to present the frequency in the form
\begin{equation}
\omega=\frac{2mc^2}{\hbar\eta(\theta_i)}
\left(1-2e^{-\tau}\right),
\label{eq42}
\end{equation}
\noindent
where the limiting case (\ref{eq41}) corresponds to $\tau\to \infty$.

Substituting Eq.~(\ref{eq42}) in Eq.~(\ref{eq32}), one finds that at
$\tau\to \infty$
\begin{equation}
\Phi(\omega,\theta_i)\approx 4mc(1-\tau)\approx -4mc\tau.
\label{eq43}
\end{equation}
\noindent
Then from Eq.~(\ref{eq33}) we conclude that in the limiting case (\ref{eq41})
\begin{equation}
{\cal R}_{\rm TM}(\omega,\theta_i)\to 1, \quad{\cal R}_{\rm TE}(\omega,\theta_i)
\to 1.
\label{eq44}
\end{equation}

The next region to consider is given by Eq.~(\ref{eq26}). Here, the second line
of Eq.~(\ref{eq7}) is applicable, where the function $\Phi$ has both real and
imaginary parts
\begin{eqnarray}
&&
{\rm Re}\Phi(\omega,\theta_i)=
4\left[mc-\frac{\hbar\omega\eta}{2c}\left(1+
\frac{4m^2c^4}{\hbar^2\omega^2\eta^2}\right)
{\rm arctanh}\frac{2mc^2}{\hbar\omega\eta}\right],
\nonumber \\[1mm]
&&
{\rm Im}\Phi(\omega,\theta_i)=-\frac{\pi\hbar\omega\eta}{c}\left(1+
\frac{4m^2c^4}{\hbar^2\omega^2\eta^2}\right).
\label{eq45}
\end{eqnarray}

Using Eqs.~(\ref{eq2}) and (\ref{eq6}), we find the reflectivities in the form
\begin{eqnarray}
&&
{\cal R}_{\rm TM}(\omega,\theta_i)=
\frac{\alpha^2\cos^2(\theta_i)[{\rm Re}^2\Phi(\omega,\theta_i)+
{\rm Im}^2\Phi(\omega,\theta_i)]}{\alpha^2\cos^2(\theta_i){\rm Re}^2\Phi(\omega,\theta_i)+
\left[2\hbar\frac{\omega}{c}\eta^2-\alpha\cos\theta_i
{\rm Im}\Phi(\omega,\theta_i)\right]^2},
\nonumber \\[1mm]
&&
{\cal R}_{\rm TE}(\omega,\theta_i)=
\frac{\alpha^2[{\rm Re}^2\Phi(\omega,\theta_i)+
{\rm Im}^2\Phi(\omega,\theta_i)]}{\alpha^2{\rm Re}^2\Phi(\omega,\theta_i)+
\left[2\hbar\frac{\omega}{c}\cos\theta_i-\alpha
{\rm Im}\Phi(\omega,\theta_i)\right]^2}.
\label{eq46}
\end{eqnarray}

Now we consider the asymptotic behavior of the reflectivities at high frequencies
\begin{equation}
\frac{\hbar\omega\eta(\theta_i)}{2mc^2}\gg 1.
\label{eq47}
\end{equation}
\noindent
In this case Eq.~(\ref{eq45}) leads to
\begin{eqnarray}
&&
{\rm Re}\Phi(\omega,\theta_i)\approx
-\frac{64m^3c^5}{3\hbar^2\omega^2\eta^2(\theta_i)}\to 0{\ \ }
\mbox{when}{\ }\omega\to\infty,
\nonumber \\[1mm]
&&
{\rm Im}\Phi(\omega,\theta_i)\approx -\pi\hbar\frac{\omega}{c}\eta(\theta_i),
\label{eq48}
\end{eqnarray}
i.e., to the same result as for a gapless graphene [see Eq.~(\ref{eq8})]. Then,
the reflectrivities are again given by Eq.~(\ref{eq40}) or, neglecting by the
small terms, as compared to unity, by
\begin{equation}
{\cal R}_{\rm TM}(\theta_i)\approx
\frac{\alpha^2\pi^2\cos^2\theta_i}{(2+\alpha\pi\cos\theta_i)^2},
\qquad
{\cal R}_{\rm TE}(\theta_i)\approx\frac{\alpha^2\pi^2}{(2\cos\theta_i+\alpha\pi)^2}.
\label{eq49}
\end{equation}
\noindent
Note that these reflectivities do not depend on the frequency and on the mass-gap
parameter.

The remaining asymptotic region is given by Eq.~(\ref{eq41}) when the quantity
on the left-hand side of Eq.~(\ref{eq47}) approaches unity from the right. In
this case from the second line of Eq.~(\ref{eq45}) one obtains
\begin{equation}
{\rm Im}\Phi(\omega,\theta_i)\approx -2\pi\hbar\frac{\omega}{c}\eta(\theta_i).
\label{eq50}
\end{equation}
\noindent
The asymptotic behavior of the real part of $\Phi$ in the first line of
Eq.~(\ref{eq45}) can be found similar to Eqs.~(\ref{eq42}) and (\ref{eq43})
with the same results for the reflectivity properties as in Eq.~(\ref{eq44}).

\subsection{Numerical computations}

We are coming to numerical computations of the reflectivity properties of gapped
graphene at zero temperature over wide frequency range of incident electromagnetic
waves. Computations are performed by using Eqs.~(\ref{eq2}), (\ref{eq6}) and
(\ref{eq7}). Taking into account that due to the conditions $\tilde{v}_F\ll 1$,
$\eta\approx 1$ the angular dependences of reflectivities are similar to those
for gapless graphene \cite{38}, computations are performed at the normal incidence.
In Fig.~1 the three lines from right to left show the
computational results for the TM and TE reflectivities (\ref{eq34}) at the normal
incidence calculated as functions of frequency for the graphene sheets with
mass-gap parameter $m$ equal to 0.1, 0.05, and 0.01\,eV, respectively. The double
logarithmic scale is used allowing to cover the frequency range from 1\,$\mu$eV to
20\,eV. The linear dependence of the logarithm of reflectivities on the
logarithm of frequency at low frequencies satisfying the condition (\ref{eq35})
corresponds to the asymptotic regime where Eq.~(\ref{eq37}) is applicable. At the
frequency value
\begin{equation}
\omega=\frac{2mc^2}{\hbar},
\label{eq51}
\end{equation}
\noindent
which is different for different $m$ [recall that at the normal
incidence $\eta(0)=1$], the narrow
resonances are observed in Fig.~1. Under the condition (\ref{eq41}) these
resonances satisfy Eq.~(\ref{eq44}). At high frequencies satisfying the
condition (\ref{eq47}) the asymptotic equations (\ref{eq49}) are applicable.
Note that at $\theta_i=0$ one can neglect by $\alpha\pi$, as compared with 2
in the denominators.

It would be interesting to investigate in more detail an immediate vicinity
of the resonance frequency (\ref{eq51}), where the above asymptotic expressions
do not apply. As an example, in Fig.~2 we again plot the reflectivity at the
normal incidence (\ref{eq34}) as a function of frequency for the graphene sheet
with $m = 0.1\,eV$. Now the computational results are plotted over a narrow
frequency region from 0.14 to 0.30\,eV using the natural scale in the frequency
axis. As is seen in Fig.~2, even over a so narrow range, the reflectivity of
graphene varies by four orders of magnitude, whereas the width of the resonance
peak remains unresolved.

In fact the half-width of this resonance can be found analytically from the
asymptotic representation (\ref{eq43}). Substituting Eq.~(\ref{eq43}) in
Eq.~(\ref{eq33}) taken at $\theta_i=0$, we find
\begin{equation}
{\cal R}_{\rm TM}(\omega,0)=
{\cal R}_{\rm TM}(\omega,0)\approx
\frac{\alpha^2\tau^2}{\frac{\hbar^2\omega^2}{4m^2c^4}+\alpha^2\tau^2}.
\label{eq52}
\end{equation}
\noindent
Taking into account Eq.~(\ref{eq51}), one obtains that
${\cal R}_{\rm TM\,(TE)} \approx 1/2$ when $\alpha^2\tau^2 = 1$, i.e., for
$\tau=1/\alpha \approx 137$. Then, using Eq.~(\ref{eq42}), we conclude that
the half-width of the resonance is equal to
\begin{equation}
4e^{-137}\frac{mc^2}{\hbar}\sim 10^{-59}\frac{mc^2}{\hbar}.
\label{eq53}
\end{equation}
\noindent
This half-width is too small and makes the resonance under consideration
unobservable. However, at zero temperature it might be possible to observe the
change in the reflectivity of graphene by a factor of ten with
increasing frequency of the incident light. Thus, the reflectivity increases
from 0.00013 to 0.00134 when $\omega$ increases from 0.16 to 0.199\,eV. In a
similar way, with further increase of frequency from 0.201 to 0.3\,eV the
reflectivity of graphene changes from 0.00141 to 0.000141.

We consider now more detailly the region of frequencies from 0 to 0.4\,eV,
where the reflectivity varies by more than an order of magnitude for all
values of $m$ under discussion. Note that at room temperature ($T=300\,$K)
it holds $k_{B}T \approx 0.025$\,eV, i.e., the thermal energy is of the same
order as $mc^2$ for the smallest value of $m$.

In Fig.~3 we plot the reflectivity of gapped graphene at the normal incidence
as a function of frequency by the three lines from right to left for the
mass-gap parameter $m=0.1$, 0.05, and 0.01\,eV, respectively. In this figure,
both the reflectivity and frequency are plotted in their natural scales.
Because of this, only the lower parts of the resonance peaks are shown.
As is seen in Fig.~3, an impact of the mass-gap parameter of graphene on
the reflectivity properties decreases with decreasing $m$ and increasing
frequency. For small mass-gap parameters $m<0.01$\,eV in the region
$\hbar\omega>2mc^2$ the reflectivity is approximately equal to that obtained
for $m=0$ at any temperature $T \leq 300$\,K \cite{38}. The role of nonzero
temperature in the case of gapped graphene is considered below.

\section{Influence of nonzero temperature on the reflectivity of gapped graphene}

In this section, we consider the reflection coefficients (\ref{eq2}), where
the quantities $\Pi_{00}$ and $\Pi$ are given by Eq.~(\ref{eq5}), i.e., are
defined at any nonzero temperature. In so doing, the zero-temperature
contributions are presented in Eqs.~(\ref{eq6}) and (\ref{eq7}) and
the thermal corrections to them are given in Eqs.~(\ref{eq16}) and (\ref{eq30})
with all necessary notations in Eqs.~(\ref{eq17})--(\ref{eq21}), (\ref{eq25}),
 (\ref{eq27}) and (\ref{eq31}).

\subsection{Asymptotic results}

We begin with thermal correction to the polarization tensor of graphene
calculated at frequencies satisfying Eq.~(\ref{eq24}). From Eqs.~(\ref{eq16}),
(\ref{eq17}), (\ref{eq20}), (\ref{eq21}) and (\ref{eq25}) one obtains
\begin{eqnarray}
&&
\Delta_T\Pi_{00}(\omega,\theta_i)=\frac{16\alpha\hbar}{\tilde{v}_F^2}
\int_{mc/\hbar}^{\infty}\frac{du}{e^{\beta u}+1}
\left\{
\vphantom{\left[\frac{(2cu-\omega)^2-\tilde{v}_F^2\omega^2
\sin^2\theta_i}{\sqrt{(2cu-\omega)^2+\tilde{v}_F^2\omega^2
\left(\frac{4m^2c^4}{\hbar^2\omega^2\eta^2}-1\right)
\sin^2\theta_i}}\right]}
1+\frac{1}{2\omega\eta}
\right.
\nonumber \\
&&
\times\left[-
\frac{(2cu+\omega)^2-\tilde{v}_F^2\omega^2
\sin^2\theta_i}{\sqrt{(2cu+\omega)^2+\tilde{v}_F^2\omega^2
\left(\frac{4m^2c^4}{\hbar^2\omega^2\eta^2}-1\right)
\sin^2\theta_i}}\right.
\nonumber \\[-2mm]
&&\label{eq54} \\[-0.3mm]
&&
+\left.\left.
\frac{(2cu-\omega)^2-\tilde{v}_F^2\omega^2
\sin^2\theta_i}{\sqrt{(2cu-\omega)^2+\tilde{v}_F^2\omega^2
\left(\frac{4m^2c^4}{\hbar^2\omega^2\eta^2}-1\right)
\sin^2\theta_i}}\right]\right\}.
\nonumber
\end{eqnarray}

Now we impose a more stringent than Eq.~(\ref{eq24}) requirement, i.e.,
$\hbar\omega\lesssim mc^2$.Under this requirement the quantity $2cu-\omega$
cannot become too small and one can expand both square roots in Eq.~(\ref{eq54})
up to the first order of respective small parameters with the result
\begin{eqnarray}
&&
\Delta_T\Pi_{00}(\omega,\theta_i)\approx\frac{16\alpha\hbar}{\tilde{v}_F^2}
\int_{mc/\hbar}^{\infty}\frac{du}{e^{\beta u}+1}
\label{eq55} \\
&&~~
\times\left\{ 1+\frac{1}{2\eta}\left[-2+
\frac{\tilde{v}_F^2\sin^2\theta_i}{2}
\left(\frac{4m^2c^4}{\hbar^2\omega^2\eta^2}+1\right)
\left(
\frac{\omega}{2cu+\omega}-\frac{\omega}{2cu-\omega}\right)
\right]\right\}.
\nonumber
\end{eqnarray}
\noindent
Then, using the smallness of $\tilde{v}_F$, we arrive at
\begin{equation}
\Delta_T\Pi_{00}(\omega,\theta_i)\approx -8\alpha\hbar\sin^2\theta_i
 \int_{mc/\hbar}^{\infty}\frac{du}{e^{\beta u}+1}\left(1+
\frac{4\frac{m^2c^4}{\hbar^2}+\omega^2}{4u^2c^2-\omega^2}\right).
\label{eq56}
\end{equation}
\noindent
Note that Eq.~(\ref{eq56}) is also valid under a less stringent condition
(\ref{eq24}) if one assumes that $\hbar\omega\ll k_BT$. This is because the
main contribution to the integral in Eq.~(\ref{eq54}) is given by
$\hbar cu\sim k_BT$ and hence the quantity $2cu-\omega$ cannot become too small.
Both application regions of Eq.~(\ref{eq56}) are used below to obtain approximate
expressions for the polarization tensor and reflectivity of gapped graphene at
nonzero temperature.

It is convenient to introduce two dimensionless parameters and new integration
variable according to
\begin{equation}
\mu\equiv\frac{mc^2}{k_BT}, \quad
\nu\equiv\frac{\hbar\omega}{k_BT}, \quad
v=\frac{\hbar cu}{k_BT}\equiv\beta u.
\label{eq57}
\end{equation}
\noindent
Then Eq.~(\ref{eq56}) takes the form
\begin{equation}
\Delta_T\Pi_{00}(\omega,\theta_i)\approx -8\alpha\sin^2\theta_i
\frac{k_BT}{c}
\int_{\mu}^{\infty}\frac{dv}{e^{v}+1}\left(1+
\frac{4\mu^2+\nu^2}{4v^2-\nu^2}\right).
\label{eq58}
\end{equation}

We consider, first, Eq.~(\ref{eq58}) under its validity condition $\nu\ll 1$
and assume that $\mu\ll 1$ as well.
Taking into account that the main contribution to the integral in Eq.~(\ref{eq58})
is given by $v\sim 1$, one can neglect by the small second term in the round
brackets. Then, after the integration, we arrive at
\begin{equation}
\Delta_T\Pi_{00}(\omega,\theta_i)\approx -8\alpha
\frac{k_BT}{c}\sin^2\theta_i\ln\left(1+e^{-\mu}\right)
\approx -8\alpha\ln 2
\frac{k_BT}{c}\sin^2\theta_i.
\label{eq59}
\end{equation}

The latter thermal correction does not depend on $m$ and coincides with that
obtained in Eq.~(89) of Ref.~\cite{38} for a gapless graphene. Note, however,
that for a gapped graphene at $T=0\,$K from Eqs.~(\ref{eq6}) and (\ref{eq36}),
i.e., in the case of small frequencies, one obtains
\begin{equation}
\Pi_{00}^{(0)}(\omega,\theta_i)\approx \frac{4}{3}\alpha\sin^2\theta_i
\frac{\hbar^2\omega^2}{mc^3}.
\label{eq60}
\end{equation}
\noindent
This is real quantity, as opposed to the case of $m=0$, where $\Pi_{00}^{(0)}$
at small frequencies is pure imaginary.

As is seen using Eq.~(\ref{eq24}) and the conditions $\mu,\,\nu\ll 1$,
$\Pi_{00}^{(0)}\ll\Delta_T\Pi_{00}$. As a result, for the TM reflectivity of graphene
 from Eqs.~(\ref{eq2}) and (\ref{eq59}) we find
\begin{equation}
{\cal R}_{\rm TM}(\omega,\theta_i)\approx
\frac{16\alpha^2\ln^2{2}(k_BT)^2\cos^2\theta_i}{\hbar^2\omega^2+
16\alpha^2\ln^2{2}(k_BT)^2\cos^2\theta_i}.
\label{eq61}
\end{equation}
\noindent
Similar approximate calculation with account of Eqs.~(\ref{eq6}) and (\ref{eq36})
results in
\begin{equation}
\Delta_T\Pi(\omega,\theta_i)\approx 8\alpha\ln 2\frac{\omega^2}{c^2}
\frac{k_BT}{c}\sin^2\theta_i\gg|\Pi^{(0)}(\omega,\theta_i)|.
\label{eq61a}
\end{equation}
\noindent
Then, substituting this equation in Eq.~(\ref{eq2}), one obtains
\begin{equation}
{\cal R}_{\rm TE}(\omega,\theta_i)\approx
\frac{16\alpha^2\ln^2{2}(k_BT)^2}{\hbar^2\omega^2\cos^2\theta_i+
16\alpha^2\ln^2{2}(k_BT)^2}.
\label{eq61b}
\end{equation}

Unlike the case of gapped graphene at zero temperature, Eqs.~(\ref{eq61}) and
(\ref{eq61b}) result in
\begin{equation}
{\cal R}_{\rm TM\,(TE)}(\omega,\theta_i)\to 1{\ \ }\mbox{when}{\ }{\omega\to 0}
\label{eq62}
\end{equation}
\noindent
similar to the case of gapless graphene at nonzero $T$ \cite{38}. Below we demonstrate,
both analytically and numerically, that this is general property of the reflectivity
of gapped graphene at $T\neq 0$.

Now we return to Eq.~(\ref{eq58}) derived under the condition
$\hbar\omega\lesssim mc^2$ and consider it under additional assumptions
$\mu\gg 1$ and $\hbar\omega\ll mc^2$. Under these assumptions one can neglect by
unity, as compared with $e^v$, and by $\nu^2$, as compared with $\mu^2$ and $v^2$.
As a result, Eq.~(\ref{eq58}) takes the form
\begin{eqnarray}
&&
\Delta_T\Pi_{00}(\omega,\theta_i)\approx -8\alpha\sin^2\theta_i
\frac{k_BT}{c}
\int_{\mu}^{\infty}\!\!\!{dv}{e^{-v}}\left(1+
\frac{\mu^2}{v^2}\right)
\nonumber \\[-1mm]
&&\label{eq63}\\[-0.5mm]
&&~
=-8\alpha\sin^2\theta_i\frac{k_BT}{c}\left\{e^{-\mu}+\mu^2\left[
\frac{e^{-\mu}}{\mu}+{\rm Ei}(-e^{-\mu})\right]\right\},
\nonumber
\end{eqnarray}
\noindent
where ${\rm Ei}(z)$ is the exponential integral. Taking into account the
asymptotic expansion \cite{47}
\begin{equation}
{\rm Ei}(-x)=e^{-x}\left[-\frac{1}{x}+\frac{1}{x^2}+
O\left(\frac{1}{x^3}\right)\right],
\label{eq64}
\end{equation}
\noindent
one arrives at
\begin{equation}
\Delta_T\Pi_{00}(\omega,\theta_i)\approx -16\alpha\sin^2\theta_i
\frac{k_BT}{c}e^{-mc^2/(k_BT)}.
\label{eq65}
\end{equation}
\noindent
Combining this expression with Eq.~(\ref{eq60}), we find
\begin{equation}
\Pi_{00}(\omega,\theta_i)\approx 4\alpha\sin^2\theta_i
\frac{\omega}{c}\left[\frac{1}{3}\,\frac{\hbar^2\omega}{mc^2}-
4\frac{k_BT}{\omega}e^{-mc^2/(k_BT)}\right],
\label{eq66}
\end{equation}
\noindent
i.e., the thermal correction is exponentially small.
Substituting this equation [and similar for $\Pi(\omega,\theta_i)$]
in Eq.~(\ref{eq2}), we again obtain  Eq.~(\ref{eq62})
in the limiting case of vanishing frequency.

As is seen from Eq.~(\ref{eq66}), the thermal correction to the 00 component of
the polarization tensor is negative, whereas the zero-temperature contribution
to it is positive. Taking into account that they can be of the same order of
magnitude, we find the value of frequency $\omega_0$ where the quantity $\Pi_{00}$
vanishes
\begin{equation}
\omega_0^2=12\frac{mc^2}{\hbar^2}k_BTe^{-mc^2/(k_BT)}.
\label{eq67}
\end{equation}
\noindent
At room temperature ($k_BT=0.025\,$eV) only the maximum value of the mass-gap parameter
under consideration here ($m=0.1\,$eV) leads to $\mu=4$ and can be considered as close
to the application region of the approximation used. For these values of parameters
we find from Eq.~(\ref{eq67}) the value $\omega_0=0.0234\,$eV, which is more or less in
agreement with the second application condition of our approximation
($\hbar\omega\ll mc^2$). Below we compare the obtained approximate value of $\omega_0$
with the results of numerical computations and find rather good agreement.  Just now we
only note that vanishing of $\Pi_{00}(\omega_0,\theta_i)$ leads in accordance with Eq.~(\ref{eq2})
to vanishing of the reflectivities of graphene ${\cal R}_{\rm TM\,(TE)}(\omega_0,\theta_i)$ at the
frequency $\omega_0\neq 0$. This does not happen either for a gapless graphene at any
temperature or for a gapped graphene at $T=0\,$K. In the next subsection we consider the
effect of vanishing reflectivity of graphene in more details by means of numerical
computations.

Here, we continue with the approximate analytic expressions for the polarization tensor and
consider the frequency region of higher frequencies defined in Eq.~(\ref{eq26}).
We start with the imaginary part of the thermal correction (\ref{eq16}) which arises only
for $\lambda=-1$ and is obtained from the first line of Eq.~(\ref{eq17}) and the second
line of Eq.~(\ref{eq27}):
\begin{equation}
{\rm Im}\Delta_T\Pi_{00}(\omega,\theta_i)=\frac{8\alpha\hbar}{\tilde{v}_F^2}
\int_{u^{(-)}}^{u^{(+)}}\frac{du}{e^{\beta u}+1}
\frac{\left(2cu-\omega\right)^2-\tilde{v}_F^2\omega^2
\sin^2\theta_i}{\omega\eta
\sqrt{\tilde{v}_F^2\omega^2\sin^2\theta_i\,A(\omega,\theta)-
\left(2cu-\omega\right)^2}},
\label{eq68}
\end{equation}
\noindent
where
\begin{equation}
A(\omega,\theta_i)\equiv 1-4\frac{m^2c^4}{\hbar^2\omega^2\eta^2(\theta_i)}
\label{eq69}
\end{equation}
\noindent
and $u^{(\pm)}$ are defined in Eq.~(\ref{eq23}). Note that in the case of equality
in Eq.~(\ref{eq26}) the imaginary part (\ref{eq68}) vanishes.

It is convenient to introduce new dimensionless integration variable
\begin{equation}
\tau=\frac{2cu-\omega}{\tilde{v}_F\omega\sin\theta_i\sqrt{A(\omega,\theta_i)}}.
\label{eq70}
\end{equation}
\noindent
Then Eq.~(\ref{eq68}) takes the form
\begin{equation}
{\rm Im}\Delta_T\Pi_{00}(\omega,\theta_i)=-4\alpha\hbar\frac{\omega}{c}
\frac{\sin^2\theta_i}{\eta}
\int_{-1}^{1}\frac{d\tau}{e^{\hbar\omega/(2k_BT)}
e^{\gamma\tau}+1}
\frac{1-A(\omega,\theta_i)\tau^2}{\sqrt{1-\tau^2}},
\label{eq71} 
\end{equation}
\noindent
where
\begin{equation}
\gamma\equiv
\gamma(\omega,\theta_i)= \tilde{v}_F\sin\theta_i
\sqrt{A(\omega,\theta_i)}\frac{\hbar\omega}{2k_BT}.
\label{eq72}
\end{equation}

The integral in Eq.~(\ref{eq71}) can be calculated approximately under different
relationships between $\omega$ and $T$. We first assume that $\hbar\omega\ll k_BT$
[and also use $\eta(\theta_i)\approx 1$]. In this case both exponents in the
denominator of Eq.~(\ref{eq71}) can be replaced with unities and after the integration
one obtains
\begin{equation}
{\rm Im}\Delta_T\Pi_{00}(\omega,\theta_i)\approx
-\alpha\pi\hbar\frac{\omega}{c}
\sin^2\theta_i e^{-\frac{\hbar\omega}{2k_BT}}\left(1+
\frac{4m^2c^4}{\hbar^2\omega^2}\right)
\approx
-\alpha\pi\hbar\frac{\omega}{c}
\sin^2\theta_i \left(1+\frac{4m^2c^4}{\hbar^2\omega^2}\right).
\label{eq73}
\end{equation}
\noindent
For $m=0$ this expression coincides with Eq.~(83) of Ref.~\cite{38} obtained for
gapless graphene.

Now we consider the opposite case $\hbar\omega\gg k_BT$. In this case it is convenient
to rewrite Eq.~(\ref{eq71}) in the form
\begin{equation}
{\rm Im}\Delta_T\Pi_{00}(\omega,\theta_i)=-4\alpha\hbar\frac{\omega}{c}
\frac{\sin^2\theta_i}{\eta}e^{-\frac{\hbar\omega}{2k_BT}}
\int_{-1}^{1}\frac{d\tau}{e^{\gamma\tau}+e^{-\hbar\omega/(2k_BT)}}
\frac{1-A(\omega,\theta_i)\tau^2}{\sqrt{1-\tau^2}}.
\label{eq74}
\end{equation}
\noindent
Now we can neglect by the second exponent in the denominator, as compared to the first,
put $\eta(\theta_i)\approx 1$ and calculate the integral with the result
\begin{equation}
{\rm Im}\Delta_T\Pi_{00}(\omega,\theta_i)\approx -4\alpha\hbar\frac{\omega}{c}
\sin^2\theta_ie^{-\frac{\hbar\omega}{2k_BT}}
\pi\left[\frac{I_1\left(\gamma\right)}{\gamma}
\left(1-\frac{4m^2c^4}{\hbar^2\omega^2}\right)+\frac{4m^2c^4}{\hbar^2\omega^2}
I_0\left(\gamma\right)\right],
\label{eq75}
\end{equation}
\noindent
where $I_n(z)$ is the modified Bessel function of the first kind.

If the frequency is not too large, i.e., $\hbar\omega\gg k_BT$ but
$\gamma(\omega,\theta_i)\ll 1$, one can put $I_1(\gamma)\approx\gamma/2$ and
$I_0(\gamma)\approx 1$. Substituting this in Eq.~(\ref{eq75}) one finds
\begin{equation}
{\rm Im}\Delta_T\Pi_{00}(\omega,\theta_i)=-2\alpha\pi\hbar\frac{\omega}{c}
\sin^2\theta_ie^{-\frac{\hbar\omega}{2k_BT}}
\left(1+\frac{4m^2c^4}{\hbar^2\omega^2}\right).
\label{eq76}
\end{equation}
\noindent
Note that for $m=0$ this result agrees with the estimation in Eq.~(70) of Ref.~\cite{38}
made for a gapless graphene.

We now turn our attention to the real part of the polarization tensor under the
condition (\ref{eq26}). It is given by Eq.~(\ref{eq16}), where all notations are
contained in the first line of Eq.~(\ref{eq17}) for $\lambda=\pm 1$, in Eq.~(\ref{eq20})
and in the first and third lines of Eq.~(\ref{eq27}). Using the dimensionless
quantities $\mu$ and $\nu$ and the integration variable $v$ introduced in Eq.~(\ref{eq57}),
the real part of the polarization tensor takes the form
\begin{equation}
{\rm Re}\Delta_T\Pi_{00}(\omega,\theta_i)=\frac{16\alpha}{\tilde{v}_F^2}
\frac{k_BT}{c}
(I_1+I_2+I_3),
\label{eq77}
\end{equation}
\noindent
where the integrals $I_{1,2,3}$ are defined as
\begin{eqnarray}
I_1&\equiv &\int_{\mu}^{v^{(-)}}\frac{dv}{e^v+1}\left\{1-\frac{1}{2\nu\eta}\left[
\frac{(2v+\nu)^2-\tilde{v}_F^2\nu^2\sin^2\theta_i}{\sqrt{(2v+\nu)^2-
\tilde{v}_F^2\nu^2A\sin^2\theta_i}}\right.\right.
\nonumber \\[1mm]
&&+\left.\left.
\frac{(2v-\nu)^2-\tilde{v}_F^2\nu^2\sin^2\theta_i}{\sqrt{(2v-\nu)^2-
\tilde{v}_F^2\nu^2A\sin^2\theta_i}}\right]\right\},
\nonumber \\[1.5mm]
I_2&\equiv &\int_{v^{(-)}}^{v^{(+)}}\frac{dv}{e^v+1}\left[1-\frac{1}{2\nu\eta}
\frac{(2v+\nu)^2-\tilde{v}_F^2\nu^2\sin^2\theta_i}{\sqrt{(2v+\nu)^2-
\tilde{v}_F^2\nu^2A\sin^2\theta_i}}\right],
\nonumber \\[1.5mm]
I_3&\equiv &\int_{v^{(+)}}^{\infty}\frac{dv}{e^v+1}\left\{1-\frac{1}{2\nu\eta}\left[
\frac{(2v+\nu)^2-\tilde{v}_F^2\nu^2\sin^2\theta_i}{\sqrt{(2v+\nu)^2-
\tilde{v}_F^2\nu^2A\sin^2\theta_i}}\right.\right.
\nonumber \\[1mm]
&&-\left.\left.
\frac{(2v-\nu)^2-\tilde{v}_F^2\nu^2\sin^2\theta_i}{\sqrt{(2v-\nu)^2-
\tilde{v}_F^2\nu^2A\sin^2\theta_i}}\right]\right\}.
\label{eq78}
\end{eqnarray}
\noindent
Here, the integration limits are $v^{(\pm)}\equiv\beta u^{(\pm)}$ and
$u^{(\pm)}$ are presented in Eq.~(\ref{eq23}), $\eta\equiv\eta(\theta_i)$ is
defined below Eq.~(\ref{eq6}) and $A\equiv A(\omega,\theta_i)$ is introduced
in Eq.~(\ref{eq69}).

We consider first the region of relatively low frequencies $\hbar\omega\ll k_BT$
($\nu\ll 1$). Note that under the condition (\ref{eq26}) assumed now the inequality
$\nu\ll 1$ can be valid only if $\mu\ll 1$ is valid as well. Now we take into
account that the main contributions to all integrals (\ref{eq78}) are given by
$v\sim 1$ and that in our case the inequalities $\nu,\mu\ll 1$ lead to $v^{(\pm)}\ll 1$.
Therefore only the integral $I_3$ contains the values $v\sim 1$ in the integration region
and, thus, $I_3$ alone determines the value of ${\rm Re}\Delta_T\Pi_{00}$ in
this case. Under the integral $I_3$ one can expand in powers
of a small parameter $\tilde{v}_F^2\equiv v_F^2/c^2$ because for $v\sim 1$ we
have $2v\gg\nu$ and the quantity $2v-\nu$ cannot vanish. As a result, one obtains
\begin{eqnarray}
&&
{\rm Re}\Delta_T\Pi_{00}(\omega,\theta_i)\approx
-8\alpha\frac{k_BT}{c}\sin^2\theta_i\int_{v^{(+)}}^{\infty}\frac{dv}{e^v+1}
\label{eq79} \\
&&~~=
-8\alpha\frac{k_BT}{c}\sin^2\theta_i\ln\left[1+e^{-\hbar\omega/(2k_BT)}\right]
\approx
-8\alpha\ln 2\frac{k_BT}{c}\sin^2\theta_i.
\nonumber
\end{eqnarray}
\noindent
Comparing this with Eq.~(\ref{eq59}), we conclude that under the condition
$\mu\ll 1$ ($mc^2\ll k_BT$) there is a smooth transition between the frequency regions
satisfying the inequalities (\ref{eq24}) and (\ref{eq26}).

At this point we can present the complete polarization tensor of graphene under the
conditions $mc^2\ll\hbar\omega\ll k_BT$. In this case the zero-temperature contribution
$\Pi_{00}^{(0)}$ is given by Eqs.~(\ref{eq6}) and (\ref{eq48}), whereas the imaginary and
real parts of thermal correction are found in Eqs.~(\ref{eq73}) and (\ref{eq79}).
The results is
\begin{equation}
\Pi_{00}(\omega,\theta)\approx -8\alpha\ln 2\frac{k_BT}{c}\sin^2\theta_i.
\label{eq79a}
\end{equation}
\noindent
It is seen that under our conditions the imaginary part of this expression is much smaller
than the real one and can be neglected. Taking into account the similar result for
$\Pi(\omega,\theta_i)$ and using Eq.~(\ref{eq2}), for the reflectivities of graphene we
return back to Eqs.~(\ref{eq61}) and (\ref{eq61b}) obtained above in the frequency region
$\hbar\omega\ll mc^2$. We will return to this fact below when discussing the results of
numerical computations.

 Now we consider the opposite case $\hbar\omega\gg k_BT$ , i.e., $\nu\gg 1$.
In the region of frequencies satisfying Eq.~(\ref{eq26}) this is possible for both
$mc^2> k_BT$ and $mc^2< k_BT$. Then we have $v^{(\pm)}\gg 1$ and, thus,
the major contribution to  $\Delta_T\Pi_{00}$ is given by the integral $I_1$,
where the upper integration limit $v^{(-)}$ can be replaced with infinity.
Taking into account that for $v\sim 1$ we have $\nu\gg 2v$, it is easy to expand
the function under the integral $I_1$, as was done above, and arrive at
\begin{equation}
{\rm Re}\Delta_T\Pi_{00}(\omega,\theta_i)\approx
32\alpha\frac{k_BT}{c}\sin^2\theta_i
\int_{\mu}^{\infty}\frac{dv}{e^v+1}
\left(\frac{v^2}{\nu^2}+\frac{m^2c^4}{\hbar^2\omega^2}\right).
\label{eq80} 
\end{equation}
\noindent
After the integration in Eq.~(\ref{eq80}) we have
\begin{eqnarray}
&&
{\rm Re}\Delta_T\Pi_{00}(\omega,\theta_i)\approx
64\alpha\frac{k_BT}{c}\sin^2\theta_i
\label{eq81}\\
&&~~\times
\left[\left(\frac{mc^2}{\hbar\omega}\right)^2
\ln(1+e^{-\mu}) -
\frac{mc^2k_BT}{\hbar^2\omega^2}{\rm Li}_2(-e^{-\mu})
-
\left(\frac{k_BT}{\hbar\omega}\right)^2{\rm Li}_3(-e^{-\mu})\right],
\nonumber
\end{eqnarray}
\noindent
where ${\rm Li}_n(z)$ is the polylogarithm function.

If we additionally assume that $mc^2\ll k_BT$, we can neglect
by the first and second terms on the right-hand side of Eq.~(\ref{eq81}),
as compared to the third one. Then, by putting $\exp(-\mu)\approx 1$, one
arrive at
\begin{equation}
{\rm Re}\Delta_T\Pi_{00}(\omega,\theta_i)\approx
-64\alpha\frac{(k_BT)^3}{c(\hbar\omega)^2}\sin^2\theta_i\,{\rm Li}_3(-1).
\label{eq82}
\end{equation}
\noindent
Taking into account that ${\rm Li}_3(-1)=-3\zeta(3)/4$, where $\zeta(z)$ is the Riemann zeta
function, this is in agreement with Eq.~(67) of Ref.~\cite{38} obtained for $m=0$.
Then, the complete polarization tensor and  also the reflectivities of gapped graphene
in this case are the same for a gapless one [we remind that under our conditions the
imaginary part of the thermal correction to the 00 component of the
polarization tensor is exponentially small
according to Eq.~(\ref{eq76}), whereas the dominant contributions to the
imaginary parts of $\Pi_{00}^{(0)}$
and $\Pi^{(0)}$ do not depend on $m$].

Under another additional condition, $k_BT\ll mc^2$, the main
contribution in Eq.~(\ref{eq81})
is given by the first term on the right-hand side
and the result is
\begin{equation}
{\rm Re}\Delta_T\Pi_{00}(\omega,\theta_i)\approx
96\alpha\frac{k_BTm^2c^3}{(\hbar\omega)^2}\sin^2\theta_ie^{-\frac{mc^2}{k_BT}}.
\label{eq83}
\end{equation}
\noindent
In this case both the real and imaginary parts of the thermal correction are
exponentially small [see Eq.~(\ref{eq76})], and we again return to the same reflectivities,
as were obtained for a gapless graphene in Ref.~\cite{38}.

\subsection{Numerical computations}

Now we perform numerical computations of the reflectivity of graphene at the normal incidence
($\theta_i=0$) defined via the reflection coefficients by the first equalities in
Eq.~(\ref{eq33}).  Computations are performed using Eqs.~(\ref{eq5})--(\ref{eq7}),
(\ref{eq16})--(\ref{eq21}), (\ref{eq25}) and (\ref{eq27}).
As noted in Sec.~IIIB,
under the conditions $\tilde{v}_{F}\ll 1$, $\eta\approx 1$ the angular dependences of
reflectivities are similar to the case of gapless graphene. They are presented in the
above analytic expressions and in Figs.~3 and 5 of Ref.~\cite{38}.
Taking into account the
possibility of comparison with the measurement data, we consider the temperature at
a laboratory, $T=300\,$K, and cover the maximally wide range of frequencies.

In Fig.~4 we present the computational results for the reflectivities
${\cal R}_{\rm TM}={\cal R}_{\rm TE}$ as functions of frequency by the two solid lines
1 and 2 for graphene sheets with the mass-gap parameter $m$ equal to 0.1 and 0.05\,eV,
respectively. The double logarithmic scale is used.
The dashed lines 1 and 2 reproduce from Fig.~1 the respective results
at zero temperature. As is seen in Fig.~4, the reflectivity of graphene depends heavily
on the fact that temperature is not equal to zero.

First of all note that the reflectivity at zero frequency is equal to unity rather
than zero, as it was at zero temperature (see Figs.~1 and 3). This fact was already
established analytically in the asymptotic expressions (\ref{eq61}), (\ref{eq62}) and
(\ref{eq66}) valid at small frequencies.

Second, as was noted above, in the region of sufficiently small frequencies the
thermal correction $\Delta\Pi_{00}$ in Eq.~(\ref{eq65}) is real and negative.
Taking into account that at small frequencies $\Pi_{00}^{(0)}$ is real and positive
[see Eqs.~(\ref{eq6}) and (\ref{eq32})], it is possible that $\Pi_{00}$ turns into
zero at some frequency $\omega_0$ depending on $m$ and $T$. As is noted above,
the reflection coefficients and reflectivities also vanish at the frequency $\omega_0$.
{}From Fig.~4 it is seen that ${\cal R}_{\rm TM\,(TE)}(\omega_0,\theta_i)=0$ at
$\omega_0\approx0.02159$ and 0.03033\,eV for graphene sheets with $m=0.1$ and $0.05\,$eV,
respectively. An approximate analytic expression for $\omega_0$ is obtained in
Eq.~(\ref{eq67}). As discussed above, at room temperature
it can be applied only for the largest value
$m=0.1\,$eV and even in this case it is slightly outside the validity region of the
approximation used. However, comparing the numerical (0.02159\,eV) and analytic
(0.0234\,eV) results for $\omega_0$, one can conclude that the approximate expression
(\ref{eq67}) leads to only 8.4\% error.

In the region of higher frequencies satisfying Eq.~(\ref{eq26}) for the mass-gap
parameters considered in Fig.~4 the computational results are similar to those obtained
at $T=0\,$K. Here, we observe the resonance peaks at the border frequencies and the
asymptotic regime at high frequencies, which does not depend on $\omega$.
In this region the dashed line 1 coincides with the solid line 1 ($m=0.1\,$eV), and the
dashed line 2 ($m=0.05\,$eV)  only minor deviates from the solid line 2.
One can conclude that in the range of low frequencies satisfying the condition (\ref{eq24})
the reflectivity properties of gapped graphene at zero and room temperature are quite
different. Now we perform numerical computations in order to investigate these
differences in more details.

In Fig.~5, we present the reflectivity of graphene with $m=0.1$ and 0.05\,eV at $T=300\,$K
at the normal incidence in an immediate vicinity of the frequency $\omega_0$ where it
drops to zero (the lines 1 and 2, respectively). For this purpose the natural scale is
used along the frequency axis. The dashed lines 1 and 2 show the respective results
at $T=0\,$K. As is seen in Fig.~5, near the frequencies $\omega_0$ it should be possible
to observe the change in the reflectivity of graphene by several orders of magnitude.

{}From Fig.~4 it can be seen that
with decreasing $m$ the interval between the frequency $\omega_0$,
where the reflectivity vanishes, and the border frequency $2mc^2/\hbar$, where it is
equal to unity, becomes more narrow. To illustrate this tendency, in Fig.~6 we plot the
reflectivity of graphene with $m=0.001\,$eV at $T=300\,$K as a function of frequency in
the double logarithmic scale (the solid line). {}From Fig.~6 it is seen that zero
reflectivity is achieved at the frequency $\omega_0$ which is only slightly less than
0.002\,eV. Because of this, the minimum and maximum values of the reflectivity virtually
coincide.
 For not too high frequencies satisfying the condition $\hbar\omega\ll k_BT$
and with exception of some vicinity of the border frequency $2mc^2/\hbar$, the solid
line in Fig.~6 is in a good agreement with the asymptotic expressions (\ref{eq61}) and
(\ref{eq61b}) taken at $\theta_i=0$.

In the same figure the reflectivity of graphene with $m=0.001\,$eV at zero
temperature is shown by the dashed line. In this case the reflectivity has only the maximum
value at 0.002\,eV and goes to zero with vanishing frequency. Note that all the above
results related to the region of frequencies (\ref{eq24}) are valid only for graphene
with nonzero mass-gap parameter. Specifically, the reflectivity of gapless graphene
does not vanish at any frequency \cite{38}.

Finally, we perform numerical computations of the frequency $\omega_0$, where the
reflectivity of graphene vanishes, as a function of the mass-gap parameter $m$.
The computational results in the region of $m$ from 0.001 to 0.1\,eV at $T=300\,$K
are shown in Fig.~7 by the solid line. As is seen in Fig.~7, with increasing $m$,
the frequency $\omega_0$ increases, achieves the maximum value at $m\approx 0.03\,$eV
and then gradually decreases. In the same figure, the border frequency
$\omega=2mc^2/\hbar$ is plotted by the dashed line. It is seen that the frequency
interval between the dashed and solid lines becomes narrower with decreasing $m$
(as already was concluded from Fig.~4 basing on only two values of $m$).
The dashed line remains above the solid line at all $m$ and almost coincides with it
at $m\lesssim 0.015\,$eV. This is in accordance with Fig.~6, where for $m=0.001\,$eV
the frequencies, where reflectivity of graphene takes the minimum and maximum
values, cannot be discerned.

\section{Conclusions and discussion}

In the foregoing, we have investigated the reflectivity properties of gapped graphene
at both zero and nonzero temperature in the framework of the Dirac model.
It was shown that the presence of nonzero mass-gap parameter has a profound effect on
the reflectivity of graphene. To find this effect, we have employed the polarization
tensor of graphene found in Ref.~\cite{38}, which was further adapted and simplified
for the case of gapped graphene. Specifically, it was shown that at zero temperature
the reflectivities of gapped graphene go to zero when the frequency vanishes. This is
not the case for a gapless graphene, where the reflectivities at zero temperature are
nonzero and depend only on the fine structure constant and the angle of incidence.
We have shown also that at nonzero temperature the reflectivities of gapped graphene
go to unity with vanishing frequency, as it is for a gapless graphene.

Another distinctive property of gapped graphene is that its reflectivities possess
the narrow resonances having the
maximum values equal to unity at the border frequency $\omega=2mc^2/[\hbar\eta(\theta_i)]$.
This is the case at both zero and nonzero temperature and can be observed experimentally
as a jump in the reflectivities of graphene when passing across the border frequency.
At frequencies larger than the border frequency there is only  minor influence of the
mass-gap parameter on the reflectivities of graphene.

The next remarkable property of gapped graphene at nonzero temperature is that in the
vicinity of some frequency $\omega_0$ smaller than the border frequency its reflectivities
drop to zero. This local minimum also can be observed as a jump in the reflectivities of
graphene when passing across the frequency $\omega_0$. There is no such effect for a
gapless graphene or for a gapped graphene at $T=0\,$K. It is caused by the fact that the
thermal correction to the polarization tensor takes negative values.

All the above effects have been derived and investigated analytically in different
frequency regions. As a result, the approximate analytic expressions for the reflectivities
of graphene and for the frequency $\omega_0$ have been found. Both the cases of the
normal incidence and an arbitrary incidence angle were considered. The analytic results
have been illustrated by numerical computations over a wide range of frequencies from
$1\,\mu$eV to 20\,eV. {}From the comparison between approximate analytic and computational
results, the exactness of the former was estimated.

The developed theory of the reflectivity of gapped graphene at any temperature is based
on the first principles of quantum electrodynamics. It can be used in many prospective
applications of graphene, such as optical detectors, optoelectronic switches and other
mentioned in Sec.~I. In future it would be interesting to generalize this theory
for the case of nonzero chemical potential (see Ref.~\cite{48} for the polarization tensor
of graphene taking the chemical potential into account) and to apply it to graphene-coated
substrates.

\section*{Acknowledgments}
The authors are grateful to  M.~Bordag for helpful discussions.


\newpage
\begin{figure}[b]
\vspace*{-8cm}
\centerline{\hspace*{2.5cm}
\includegraphics{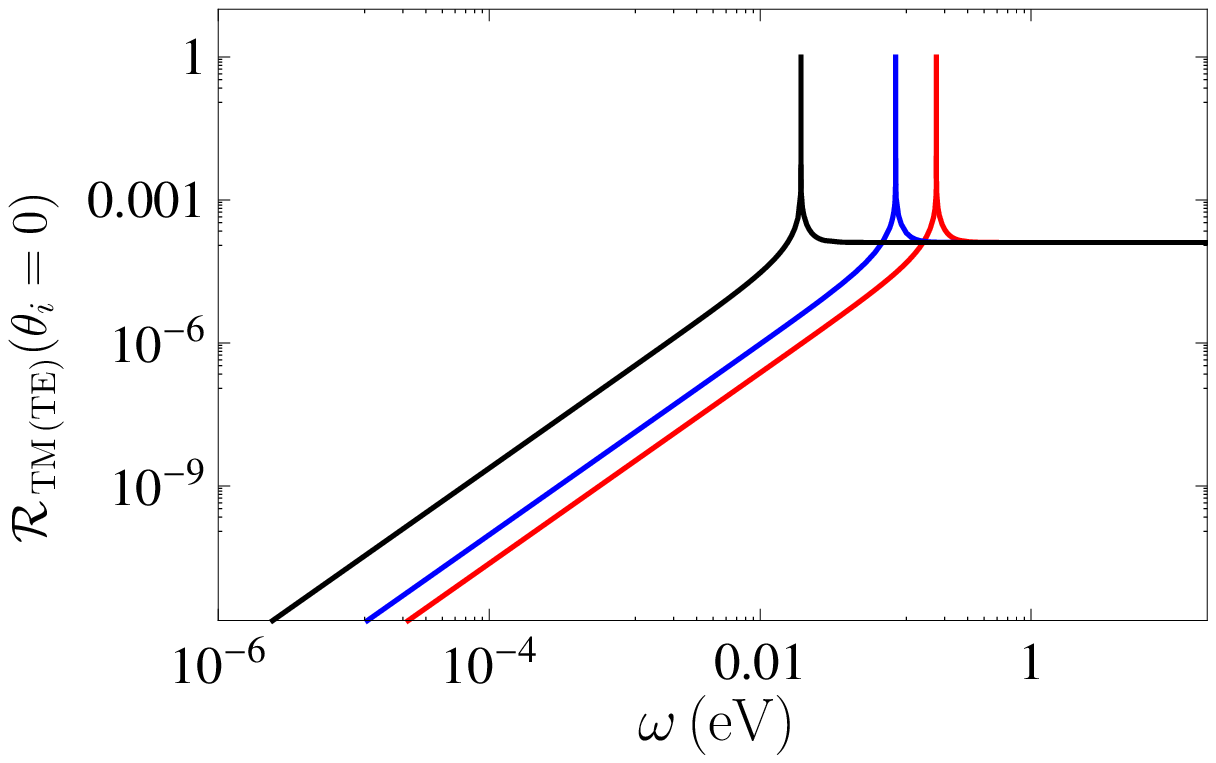}
}
\vspace*{-9cm}
\caption{\label{fg1}(Color online)
The reflectivity of gapped graphene  at $T=0\,$K
at the normal incidence is shown as a
function of frequency by the three lines from right to left for the
mass-gap parameter $m$=0.1, 0.05, and 0.01\,eV, respectively.
}
\end{figure}
\begin{figure}[b]
\vspace*{-8cm}
\centerline{\hspace*{2.5cm}
\includegraphics{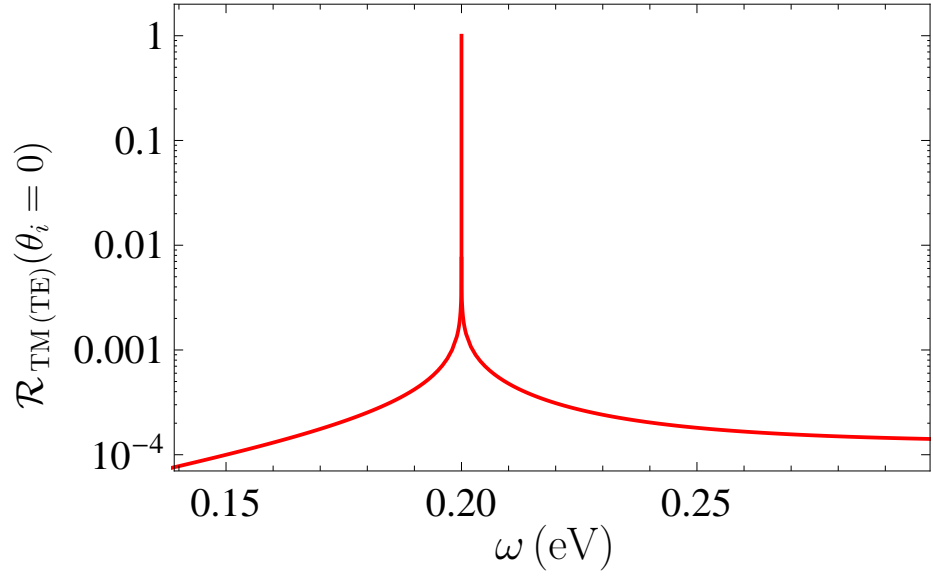}
}
\vspace*{-9cm}
\caption{\label{fg2}(Color online)
The reflectivity of graphene with the mass-gap parameter $m$=0.1\,eV
 at $T=0\,$K at the
normal incidence is shown as a function of frequency in the vicinity of the
resonance frequency $\omega=2mc^2/\hbar$.
}
\end{figure}
\begin{figure}[b]
\vspace*{-8cm}
\centerline{\hspace*{2.5cm}
\includegraphics{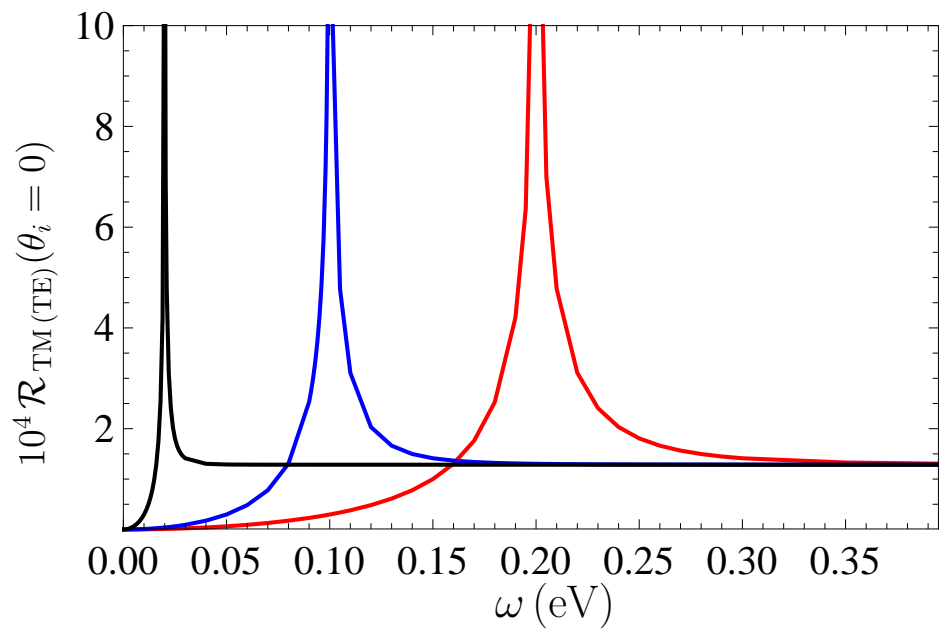}
}
\vspace*{-9cm}
\caption{\label{fg3}(Color online)
The reflectivity of gapped graphene at $T=0\,$K at
the normal incidence is shown as a
function of frequency using the natural scales in the vicinity of resonance
frequencies. The three lines from right to left are for the mass-gap parameter
$m$=0.1, 0.05, and 0.01\,eV, respectively.
}
\end{figure}
\begin{figure}[b]
\vspace*{-8cm}
\centerline{\hspace*{2.5cm}
\includegraphics{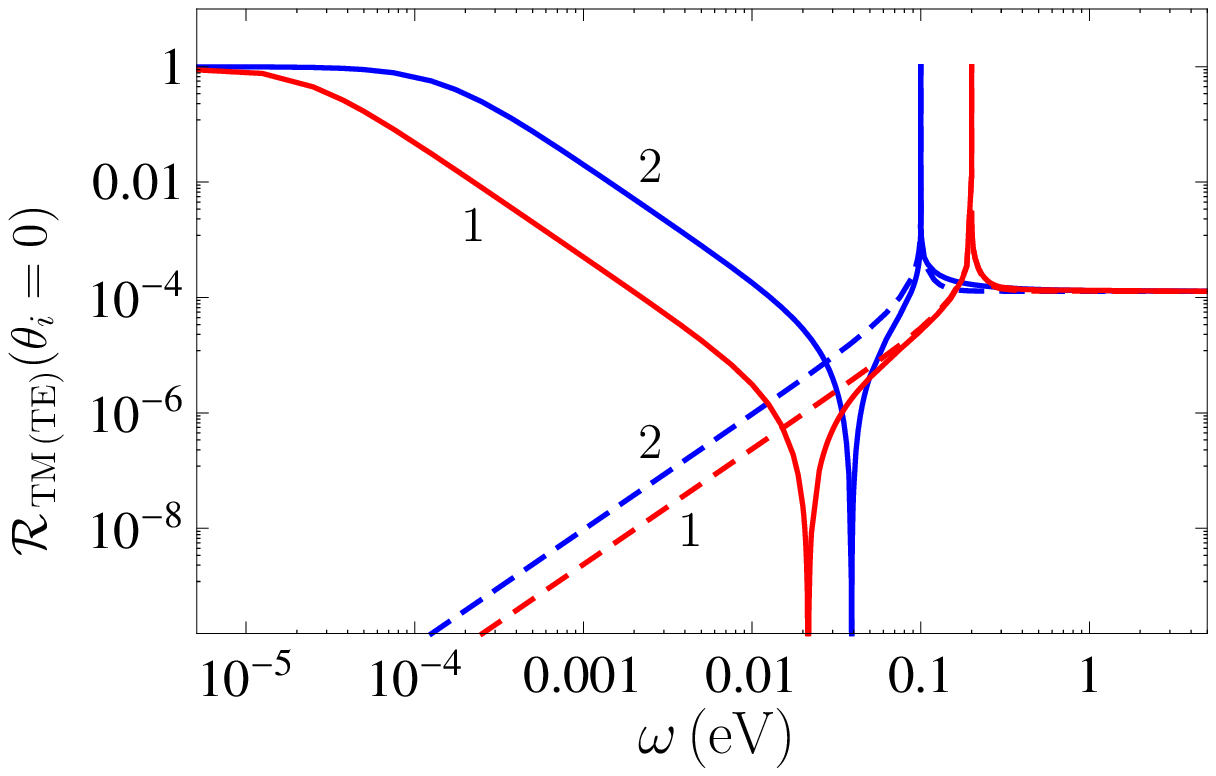}
}
\vspace*{-9cm}
\caption{\label{fg4}(Color online)
The reflectivity of gapped graphene at $T=300\,$K at the normal incidence is shown
by the solid lines 1 and 2 as a function of frequency for the mass-gap parameters
$m=0.1$ and 0.05\,eV, respectively. The dashed lines 1 and 2 reproduce the respective
results at $T=0\,$K.
}
\end{figure}
\begin{figure}[b]
\vspace*{-8cm}
\centerline{\hspace*{2.5cm}
\includegraphics{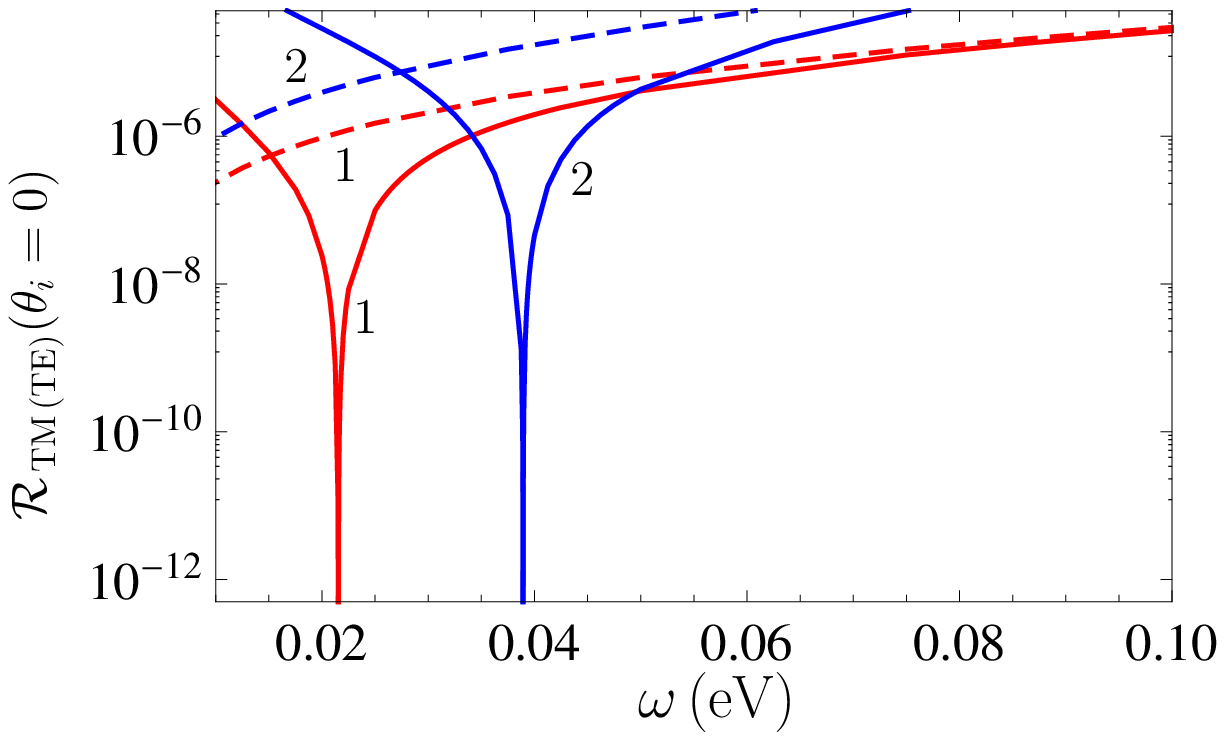}
}
\vspace*{-9cm}
\caption{\label{fg5}(Color online)
The reflectivity of gapped graphene at $T=300\,$K at the normal incidence is shown
in the vicinity of frequencies where it drops to zero
by the solid lines 1 and 2 as a function of frequency for the mass-gap parameters
$m=0.1$ and 0.05\,eV, respectively. The dashed lines 1 and 2 reproduce the respective
results at $T=0\,$K.
}
\end{figure}
\begin{figure}[b]
\vspace*{-8cm}
\centerline{\hspace*{2.5cm}
\includegraphics{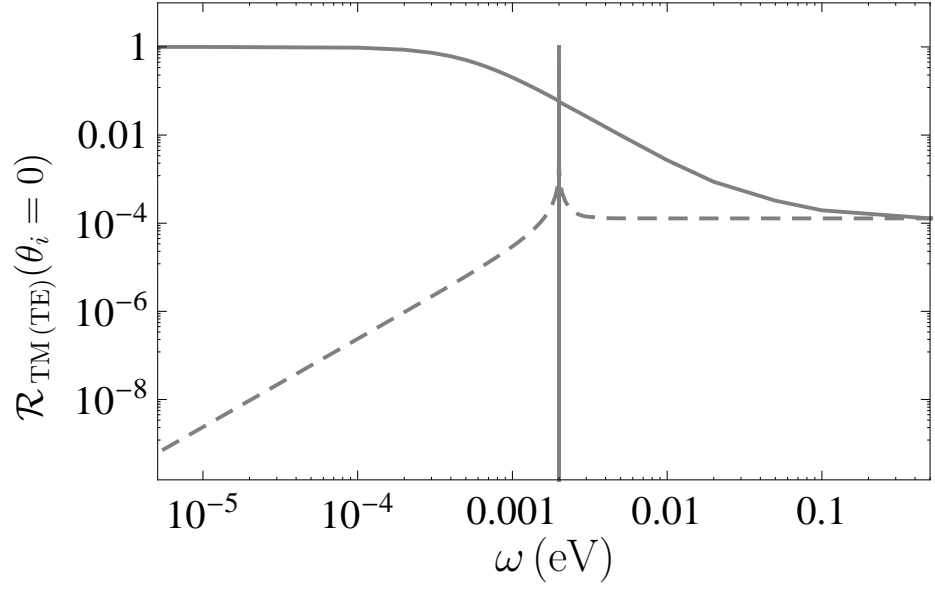}
}
\vspace*{-9cm}
\caption{\label{fg4}
The reflectivity of gapped graphene  with $m=0.001$\,eV
at the normal incidence is shown as a function of frequency
by the solid and dashed lines at $T=300\,$K and $T=0\,$K,
respectively.
}
\end{figure}
\begin{figure}[b]
\vspace*{-8cm}
\centerline{\hspace*{2.5cm}
\includegraphics{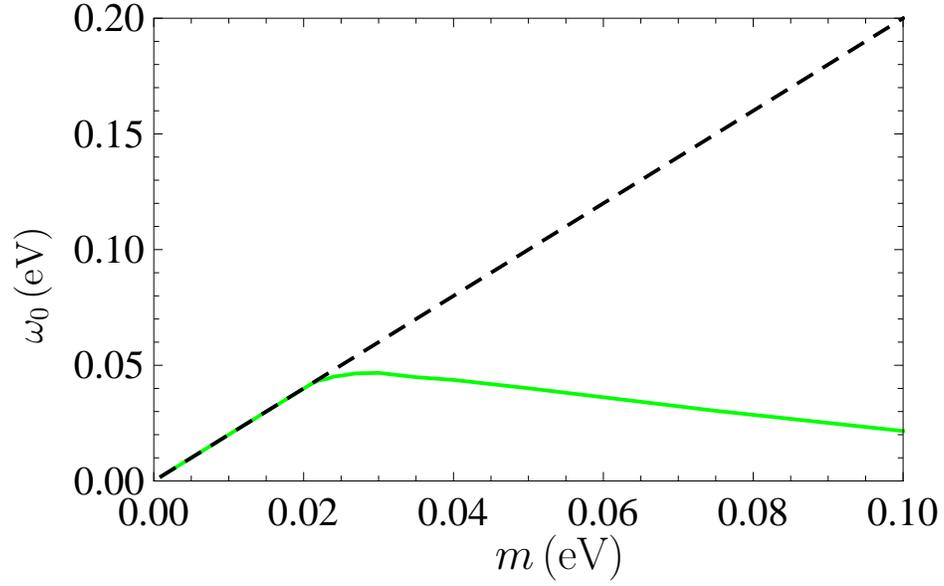}
}
\vspace*{-9cm}
\caption{\label{fg4}(Color online)
The frequency, where the reflectivity of graphene at $T=300\,$K
at the normal incidence vanishes, is shown
by the solid line as a function of the mass-gap parameter.
 The dashed line shows the border frequency $2mc^2/\hbar$.
}
\end{figure}
\end{document}